\documentclass[aps,prl,onecolumn,showpacs,preprint]{revtex4}

\usepackage{graphicx,amssymb,amsfonts,amsmath,color}
\usepackage{hyperref}
\usepackage{subfigure}

\begin{document}

\title{Generation of stable overlaps between antiparallel filaments}

\author{D. Johann, D. Goswami and K. Kruse}

\affiliation{Theoretische Physik, Universit\"at des Saarlandes, Postfach 151150, 66041 Saarbr\"ucken, Germany}

\date{\today}

\begin{abstract}
During cell division, sister chromatids are segregated by the mitotic spindle, a bipolar assembly of interdigitating antiparallel 
polar filaments called microtubules. Establishing a stable overlap region is essential for maintenance of bipolarity, but the 
underlying mechanisms 
are poorly understood. Using a particle-based stochastic model, we find that the interplay of motors and passive cross linkers 
can robustly generate partial overlaps between antiparallel filaments. Our analysis shows that motors reduce the 
overlap in a length-dependent manner, whereas passive cross linkers increase it independently of the length. In addition to 
maintaining structural integrity, passive cross linkers can thus also have a dynamic role for size regulation.
\end{abstract}
\maketitle

Eukaryotic cells store their chromosomes in a nucleus. During nuclear division, called mitosis, the duplicated 
chromosomes segregate. This process relies on the mitotic spindle, a bipolar structure of interdigitating 
microtubules~\cite{walczak08}.
These filamentous polymers are physically cross linked by specific proteins to maintain the spindle's integrity.
Some of these cross linkers are molecular motors that can generate mechanical stresses from the hydrolysis 
of ATP, which in turn can result in relative sliding between microtubules. Whereas the gross architecture of spindles 
is conserved from yeast to human, their detailed internal organisation varies largely between species.

In spite of its vital importance, the physical principles underlying formation and size regulation of mitotic spindles 
are largely unknown. Some works emphasize the inherent ability of microtubules and molecular motors to self-organize
into spindle structures by means of their mechanical interactions and microtubule 
assembly~\cite{goshima05,rubinstein09,bouck08}. Microtubules turn over on a time scale of tens of seconds with
kinetic differences at their two ends, denoted as plus and minus. Other authors have
emphasized the role of external factors in spindle assembly and maintenance, for example, through macroscopic gradients 
of various regulatory proteins~\cite{karsenti01,fuller08,cimini06,stumpff08,greenan10}. 

\begin{figure}[h]
\centering
\includegraphics[width = 3 in]{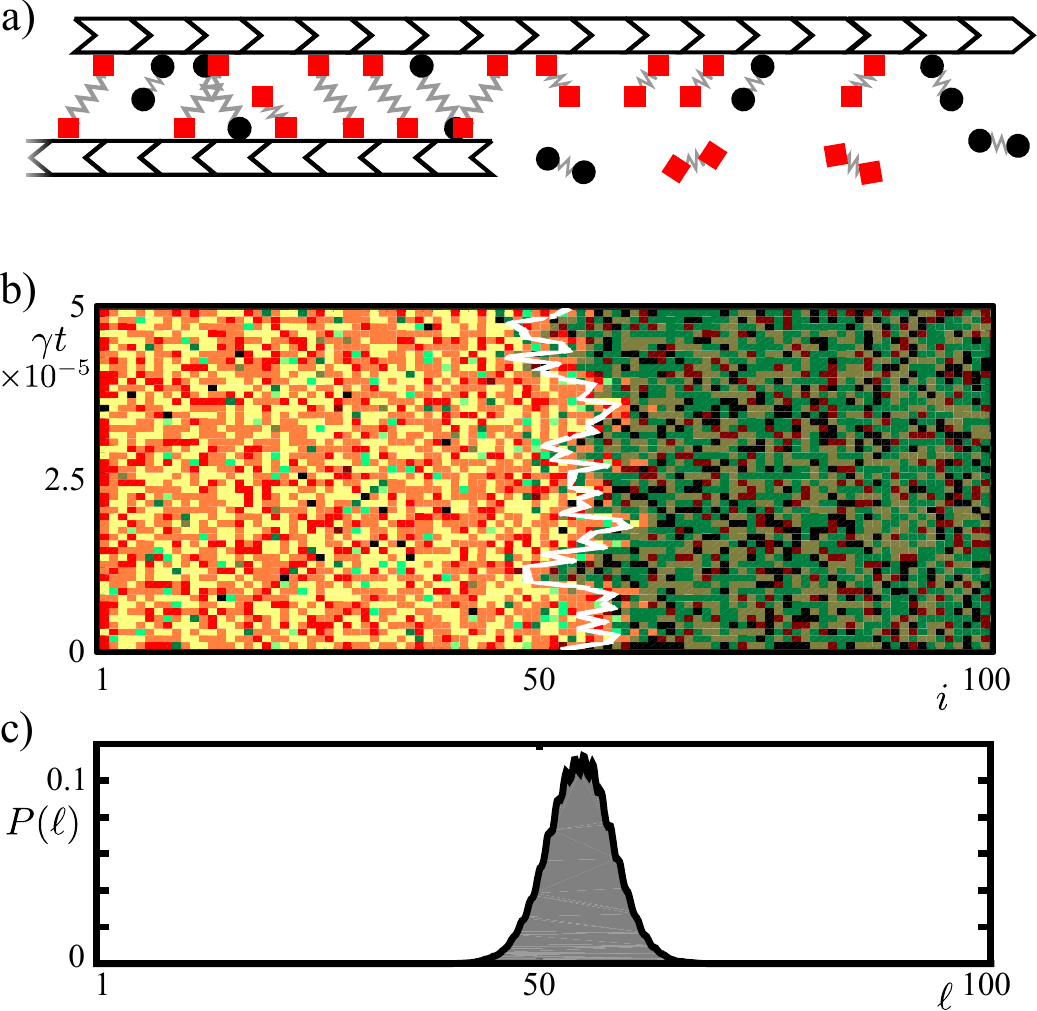}
\caption{(Color online) Stochastic model for the dynamics of antiparallel microtubules in the presence of molecular motors
and passive cross linkers. a) Schematic representation of single and double bound motors (black circles) and passive
cross linkers (red squares). b) Kymograph for the motor and cross-linker distributions on one of the filaments with its plus end 
at $i=1$. Empty sites are black, sites occupied by motors in green, and sites occupied by passive cross linkers in red. Darker colors 
indicate single bound, lighterr colors double bound particles. The white line indicates the position of the other filament's 
plus end. c) Distribution of the overlap length for the simulation in (b). Simulations are for $\omega_\mathrm{m}^\mathrm{o}=0.316$ 
and $\omega_\mathrm{p}^\mathrm{o}=10^{-4}$. For parameter values see text.}
\label{fig:model.kymo.length}
\end{figure}

On the scale of individual filaments, 
notably the regulation of microtubule length by length-dependent 
depolymerization~\cite{varga06,howard07,varga09,johann12,melbinger12} was studied. 
Less is known about mechanisms regulating the overlap between interdigitating microtubules, which is key for
maintaining spindle bipolarity in many organisms. Cross linking molecular 
motors moving towards the microtubules' plus ends tend to reduce the overlap region, see Fig.~\ref{fig:model.kymo.length}a. 
This might be compensated either by filament growth at the plus end~\cite{masuda87} or by the action of antagonistic motors 
moving towards the minus end~\cite{nedelec02}. Both mechanisms, though, turn out to require fine tuning of parameters. 
Passive cross 
linkers, that is, without motor activity, were long thought to merely provide structural integrity and to effectively increase the friction 
between sliding microtubules~\cite{tawada91}. \textit{In vitro} experiments, however, have indicated a regulatory function for the 
passive
cross linkers MAP65/PRC1/Ase1 through recruitment of specific motor proteins~\cite{bieling10} or through generating friction that 
depends on the overlap size~\cite{braun11}. Theoretical work showed that these cross linkers and molecular motors can segregate 
along microtubules~\cite{johann14}. Passive cross linkers were also shown to lead to an increase of the overlap between 
filaments~\cite{walcott10,lansky15}. Consequently, passive cross linkers might play a 
dynamically much more interesting role for spindle homeostasis than previously assumed.

In this work we show that passive cross linkers can oppose the action of molecular motors to separate two antiparallel 
microtubules resulting in the formation of a stable partial overlap, see Fig.~\ref{fig:model.kymo.length}. The 
overlap length can be tuned by varying the relative concentrations of motors and passive cross
linkers. A mean field analysis shows that the overlap results from a motor-induced sliding stress that decreases 
with decreasing overlap and an antagonistic sliding stress induced by the passive cross linkers. The latter is
independent of the overlap length and follows from an asymmetry 
in cross-linker binding at microtubule ends. Finally, we investigate the effects of steric interactions between motors and passive
cross linkers.

We describe a pair of antiparallel microtubules by two oriented lattices each with $N$ sites, see 
Fig.~\ref{fig:model.kymo.length}a.
Motors and passive cross linkers are each incorporated as particles with two identical `heads' that are connected by a linear spring of 
respective stiffnesses $k_\mathrm{m}$ and $k_\mathrm{p}$. Each site of the lattices can accommodate at most one motor 
and one passive cross-linker head. Motivated by \textit{in vitro} experiments, we consider reservoirs of motors and passive
cross linkers, implying that receptive sites are occupied by particles
from the reservoir with respective constant rates $\omega_\mathrm{m}^\mathrm{o}$ and $\omega_\mathrm{p}^\mathrm{o}$. 
They are occupied with heads of particles already bound to the other lattice
at respective rates $\omega_\mathrm{m}^\mathrm{c}$ and $\omega_\mathrm{p}^\mathrm{c}$, where the position of
a free head is distributed normally around the attached head's position with width $\sigma$. The detachment of single bound 
particles occurs 
at rates $\omega_\mathrm{m}^\mathrm{d}$ and $\omega_\mathrm{p}^\mathrm{d}$. Assuming slip 
bonds, the detachment rates of a single head of a cross-linked particle increases exponentially with the spring extension 
$\xi$, that is,  $\omega_\mathrm{m,p}^\mathrm{d}\exp\left\{k_\mathrm{m,p}\left|\xi\right|/ f_\mathrm{m,p}\right\}$, where 
$f_\mathrm{m,p}$
are characteristic force scales for motors and passive cross linkers.

Bound heads can move along a filament. Motor heads bound to a single filament 
hop at rate $\gamma$ to the neighboring site in the direction of the lattice's plus end. Heads of passive cross linkers 
hop at rate $D$ to one of the two neighboring sites. Hopping is only possible, if the target site is not 
yet carrying a head of the same species. For cross-linked particles, the hopping rates depend on the 
spring extension $\xi$ as $\gamma \exp\left\{k_\mathrm{m} \xi/f_\mathrm{m}\right\}$ and 
$D \exp\left\{\pm k_\mathrm{p} \xi/f_\mathrm{p}\right\}$. The sign is chosen such that hops reducing the spring 
extension are more likely. As for the passive cross linker
Ase1~\cite{braun11}, we consider the microtubule ends to constitute diffusive barriers: hopping 
off the lattice is suppressed. For motor particles, the boundary sites are treated as bulk sites. 

To determine the motion of the microtubules, we compute the total force generated by the extensions of all springs of 
cross-linking particles. This force is balanced by the friction force $\mu v$, where $v$ is the filament's velocity and $\mu$ 
its friction coefficient. We neglect possible Brownian motion of the filaments. This completes the description of our model.

We performed stochastic simulations, see Fig.~\ref{fig:model.kymo.length}b,c, for parameters obtained 
\textit{in vitro} for the motor Eg5 and the passive cross linker Ase1: $\omega_\mathrm{d}^\mathrm{p}= 0.0017\mathrm{s}^{-1}$, 
$\omega_\mathrm{d}^\mathrm{m}= 1.56\mathrm{s}^{-1}$, and  $\gamma = 12\mathrm{s}^{-1}$~\cite{braun11,kapitein08,valentine06}.
For the hopping rate of passive particles we chose $D=12\mathrm{s}^{-1}$~\footnote{The actual rate is 
$D= 860\mathrm{s}^{-1}$~\cite{kapitein08}. To save computation time we chose a lower value after checking that the
results are qualitatively not affected by this reduction.}. The cross-linking rates $\omega_\mathrm{m,p}^c$ 
and the spring stiffnesses $k_\mathrm{p}$ are not known experimentally. Throughout this work, we use 
$\omega_\mathrm{p}^c=1.2\mathrm{s}^{-1}$, $\omega_\mathrm{m}^c=8.4\mathrm{s}^{-1}$, and 
$k_\mathrm{m,p}=0.11\mathrm{pN}/\mathrm{nm}$~\footnote{The actual spring stiffness for Eg5 is 
$1\mathrm{pN}/\mathrm{nm}$~\cite{valentine06a}. We chose a lower value for numerical stability of the simulations.}. For the 
characteristic forces we took the stall force 
of Eg5, $f_\mathrm{m,p}=9$pN~\cite{valentine06a}, and $\sigma$ is taken to be 8nm. Each lattice 
had 100 sites corresponding to a microtubule length of 800nm and the friction coefficient $\mu$ was 
0.1pNs/nm~\footnote{This value corresponds to a viscosity 
of the surrounding fluid of $70\mathrm{Pa}\cdot\mathrm{s}$. We checked that for the viscosity of water, 
$10^{-3}\mathrm{Pa}\cdot\mathrm{s}$, for 
which the simulations take much 
longer, the distribution of the overlap length does not change.}. The rates $\omega_\mathrm{m,p}^\mathrm{o}$ depend on the 
concentrations of 
proteins in the reservoir and will be used as control parameters. From now on, we scale all rates by $\gamma$, lengths 
by 8nm, the size of a tubulin dimer, and forces by $f_\mathrm{m}$.

\begin{figure}
\includegraphics[width = 3 in]{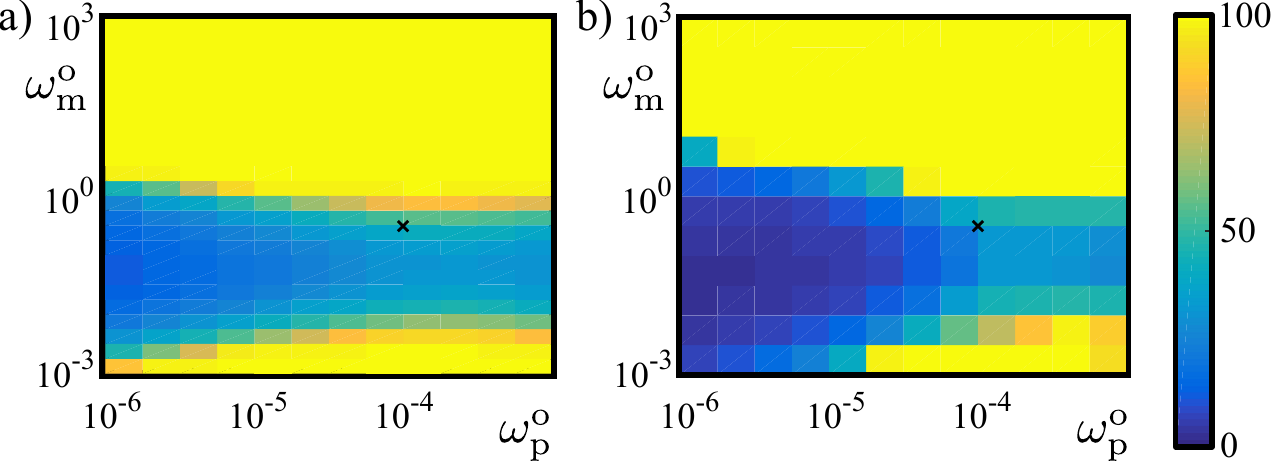}
\caption{(Color online) Overlap length as a function of the occupancy rates $\omega_\mathrm{m}^\mathrm{o}$ and
$\omega_\mathrm{p}^\mathrm{o}$ from stochastic simulations (a) and from the mean-field analysis (b). For stochastic 
simulations, the overlap length has been sampled $10^7$ times. The crosses indicate the occupancy rates used in 
Fig.~\ref{fig:model.kymo.length}b,c, other parameter values are as given in the text.}
\label{fig:phase.nosteric}
\end{figure}

For $\omega_\mathrm{m}^\mathrm{o}=0.316$ and $\omega_\mathrm{p}^\mathrm{o}=10^{-4}$, we observe a stable overlap 
between the lattices of approximately 50 sites, see Fig.~\ref{fig:model.kymo.length}b,c. The distribution of overlap lengths is 
roughly Gaussian with a standard deviation of 4 sites. The phase diagram in Fig.~\ref{fig:phase.nosteric}a shows that 
for a fixed concentration of passive particles, there is an interval of motor concentrations for which a stable partial
overlap is generated. If their concentration is below a critical value, both filaments overlap fully.
This is in line with the reports in Refs.~\cite{walcott10,lansky15} that passive cross linkers alone maximize the overlap
between two filaments. Remarkably, also for motor concentrations above a critical value, we observe full overlap.
This indicates that the ability of motors to separate antiparallel microtubules is reduced with increasing motor 
concentration. These observations are reminiscent of the dependence of the particle current in the totally asymmetric
exclusion process (TASEP), which initially increases with the particle concentration and then decreases as it 
approaches a completely filled lattice~\cite{krug91}. From Figure~\ref{fig:phase.nosteric}a, one sees that the region of 
partial overlap shrinks with the concentration of passive particles.

To understand the origin of the stable overlaps, we turned to a mean-field analysis of our system. In steady state,
the distribution of motors and passive cross linkers along the two filaments are  the same when plotted 
from the plus to the minus end for each filament. We will focus on this symmetric situation and consider the 
densities $\rho_\mathrm{m,p}(x)$ of particles bound to only one filament at a distance $x$ from the plus end. Note, that
$x$ is now a continuous variable. In addition, $\phi_\mathrm{m,p}(x,\xi)$ denote the densities of cross linking particles, 
where one head is bound at a distance $x$ 
from the plus end of one filament and $\xi$ is the spring extension. The second head is thus located a distance 
$\ell-x+\xi$ from the plus end of the other filament. 

The time evolution of $\rho_\mathrm{m,p}$ is given by the continuity equation
\begin{align}
\partial_t \rho_\mathrm{m,p} + \partial_x j_\mathrm{m,p} &= S_\mathrm{m,p}.
\label{eq:singlebound.dens}
\end{align}
Here, the currents $j_\mathrm{m,p}$ for motors and passive cross linkers are, respectively, given by
\begin{align}
j_\mathrm{m} &=\gamma \rho_\mathrm{m} (1-\rho_\mathrm{m}-Q_\mathrm{m})\\
j_\mathrm{p} &=D\left(Q_\mathrm{p} \partial_x \rho_\mathrm{p} 
	- \rho_\mathrm{p} \partial_x Q_\mathrm{p} -\partial_x \rho_\mathrm{p}\right),
\end{align}
where $Q_\mathrm{m,p} =\int d\xi\;q_\mathrm{m,p}$ with 
$q_\mathrm{m,p}=\phi_\mathrm{m,p}(x,\xi)+\phi_\mathrm{m,p}(\ell-x+\xi,\xi)$
denote the total densities of cross linking proteins at position $x$. The overlap length evolves according to
$\mu \dot\ell = F_\mathrm{m}+F_\mathrm{p}$ with
\begin{align}
F_\mathrm{m,p}=k_\mathrm{m,p}\int\limits_0^Ldx\int  
d\xi\;\xi q_\mathrm{m,p} .
\end{align} 
Finally, the source terms are given by $S_\mathrm{m,p}=\omega_\mathrm{m,p}^\mathrm{o} (1-\rho_\mathrm{m,p} -Q_\mathrm{m,p}) 
- \omega_\mathrm{m,p}^\mathrm{d} \rho_\mathrm{m,p} + \int d\xi\ 
\left[\omega_\mathrm{m,p}^\mathrm{d}\exp\left\{k_\mathrm{m,p}\left|\xi\right|/ f_\mathrm{m,p}\right\} 
q_\mathrm{m,p}- \omega_\mathrm{m,p}^\mathrm{c} 
\rho_\mathrm{m,p}(1-\bar\rho_\mathrm{m,p}-\bar Q_\mathrm{m,p})N(\xi)\right]$, where 
$N(\xi)=\exp\left\{-\xi^2/2\right\}/\sqrt{2\pi}$. Here, $\bar\rho_\mathrm{m,p}$ and $\bar Q_\mathrm{m,p}$ are the corresponding 
densities on the opposite microtubule.

The densities $\phi_\mathrm{m,p}$ of cross linking proteins also evolve according to the continuity equation. The
expressions for the currents and sources follow the same logic as for $\rho_\mathrm{m,p}$, but are rather involved, such
that we refrain from giving them here.

\begin{figure}
\includegraphics[width = 3 in]{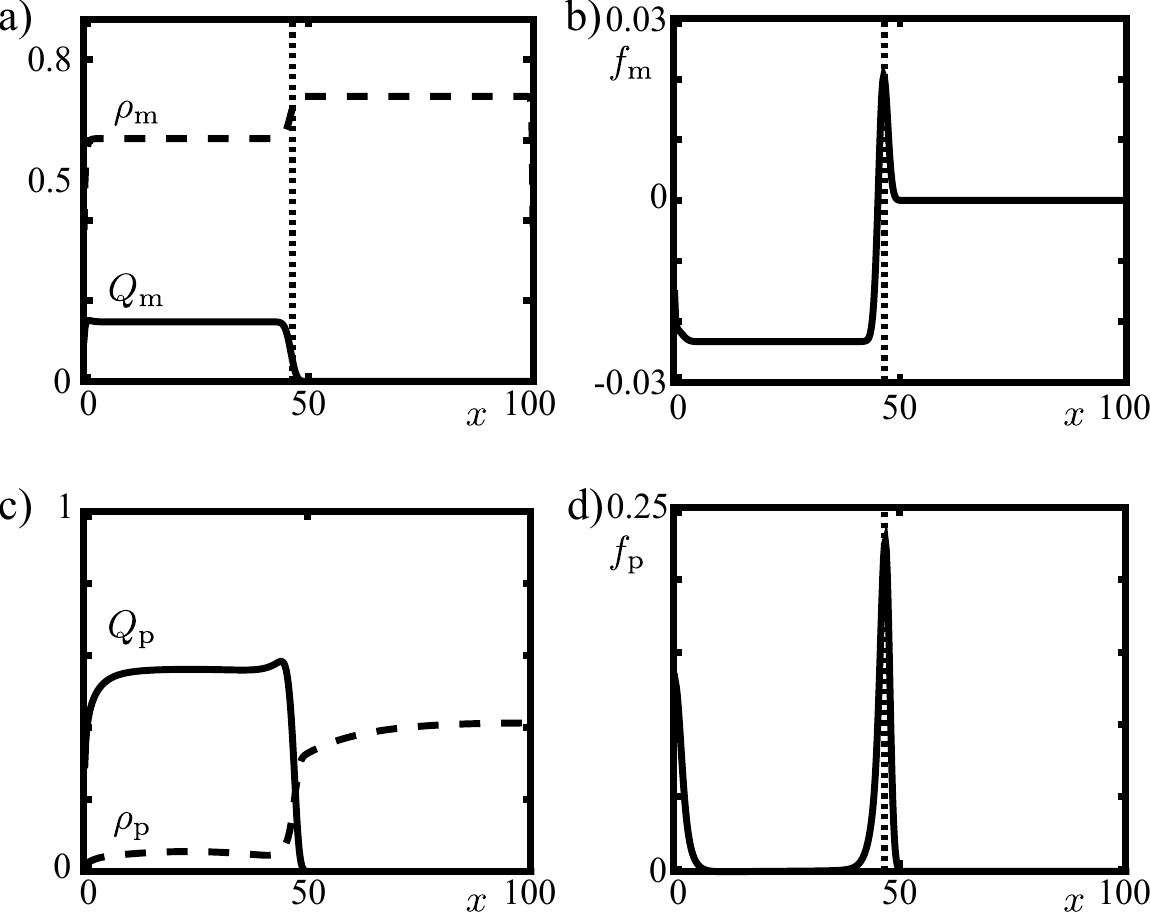}
\caption{Mean field analysis of the steady state. a,c) Densities of single bound (dashed lines) and double bound (solid lines) 
motors (a) and passive cross linkers (c). b,d) Average force density for motors (b) and passive cross linkers (d). The dotted
lines indicate the end of the overlap region. Parameters are as in Fig.~\ref{fig:model.kymo.length}b,c.}
\label{fig:meanfield.analysis}
\end{figure}

The mean-field analysis reproduces the essential features of the stochastic simulations, see Fig.~\ref{fig:phase.nosteric}b. 
The densities $\rho_\mathrm{m,p}$ and $Q_\mathrm{m,p}$ are essentially constant in the overlap and non-overlap regions of the 
filaments~\footnote{Note, that densities in the stochastic simulations fluctuate too strongly to allow for a meaningful 
comparison with the mean-field densities.}, see 
Fig.~\ref{fig:meanfield.analysis}a,c. 
The corresponding force densities are constant in the bulk of the overlap region and show pronounced maxima at 
its boundaries, see Fig.~\ref{fig:meanfield.analysis}c,d. The total force exerted by cross linking motors is dominated by the
bulk  and tends to decrease the overlap. As the overlap region shrinks, this total motor force decreases. In contrast, 
for passive cross linkers, the force density vanishes in the bulk. The
corresponding total force is given by the boundary values and independent of the overlap length, see
Fig.~\ref{fig:meanfield.analysis}d, and also tends to increase the overlap. Summing up the forces yields a stable
steady state overlap length.

At the level of individual particles, what causes the total net force of the passive cross linkers? Consider a passive cross linker 
in the bulk. The distribution of the spring 
extensions will be symmetric around zero due to the fast diffusion of bound cross linker heads and because the cross linking rate
is symmetric with respect to $\xi$. At the boundary, however, this distribution is asymmetric because binding sites are absent 
beyond the filament's plus end. This leads to a net force. 
The same effect exists for motors. However, usually the bulk forces will mask it. Only in case the motor density is so
high that motor movements are essentially blocked by steric effects does the end effect dominate. This asymmetry explains the
maximal overlaps observed for very low and very high motor densities observed in Fig.~\ref{fig:phase.nosteric}.

\begin{figure}[h]
\centering
\includegraphics[width = 3 in]{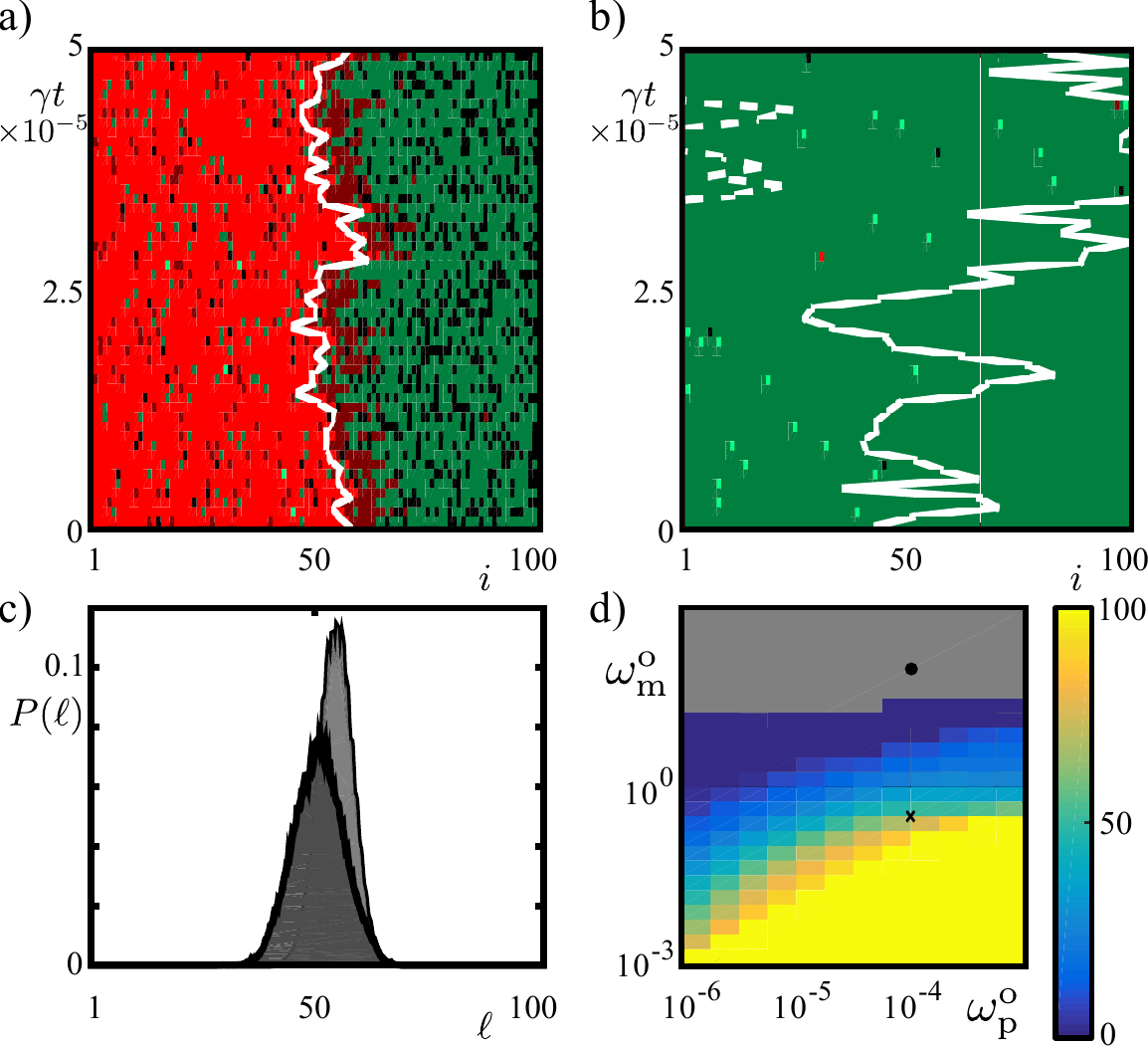}
\caption{(Color online) Stochastic model with interspecies steric interactions. a) Kymograph as in Fig.~\ref{fig:model.kymo.length}b
and for the same parameter values. b) Kymograph for $\omega_\mathrm{m}^\mathrm{o}= 10^{2}$ and 
$\omega_\mathrm{p}^\mathrm{o}$ as in (a). Dashed white line indicates the position of the other filament's minus end.
c) Distributions of overlap length for (a) (dark grey) and for Fig.~\ref{fig:model.kymo.length}c (light grey). d) Overlap 
length as a function of  $\omega_\mathrm{m}^\mathrm{o}$ and
$\omega_\mathrm{p}^\mathrm{o}$. The grey region indicates highly fluctuating states as (b). The cross indicates the 
occupancy rates used in (a), the dot those used in (b).}
\label{fig:withsteric}
\end{figure}

So far, we have not considered steric interactions between motors and passive particles. Due to their different sizes,
they might indeed occupy different protofilaments on cross linked microtubules. Still, \textit{a priori}, one cannot exclude steric 
interactions between the two protein species. Extending our model to include steric interactions, we find
that stable partial overlaps can be generated under these conditions, too, see Fig.~\ref{fig:withsteric}a. Interestingly, for a
similar average overlap length, the distribution is broader in the presence of steric 
interactions, see Fig.~\ref{fig:withsteric}c. Up to a critical
value of the occupancy rate $\omega_\mathrm{m}^\mathrm{o}$, the phase diagram looks similar to the case without
interspecies steric interactions, see Fig.~\ref{fig:withsteric}d. Above that value, we observe a phase with
fluctuations that are of the order of the filament length, see Fig.~\ref{fig:withsteric}b. More work is necessary to characterize this 
phase in detail and to identify the origin of this behavior.

In summary, our analysis shows, that in the presence of motors and passive cross linkers stable partial overlaps can be
generated between antiparallel filaments and that the overlap size can be tuned by the concentrations of motors and
passive cross linkers. Previous work has shown that, in the presence of steric interactions between 
motors, sliding can also be induced between parallel filaments~\cite{kruse02,gao15}. It will be interesting to explore 
the consequences of these collective effects for the organization of ensembles of microtubules and their possible impact on
spindle formation.

\begin{acknowledgements}
We thank M.E. Janson, G. Goshima, and S. Diez for useful discussions. 
This work was supported by the Graduate School
1276 and SFB 1027 of Deutsche Forschungsgemeinschaft and by the Human Frontiers of Science Program
Research Grant RGP0026/2011-C103.
\end{acknowledgements}


\begin{thebibliography}{999}
\bibitem{walczak08} C.E. Walczak and R. Heald, Int. Rev. Cyt.~\textbf{265}, 111 (2008).
\bibitem{goshima05} G. Goshima, R. Wollman, N. Stuurman, J.M. Scholey, and R.D. Vale, Curr Biol.~\textbf{15}, 1979  (2005).
\bibitem{rubinstein09} B. Rubinstein, K. Larripa, P. Sommi, and A. Mogilner, Phys. Biol.~\textbf{6}, 16005 (2009).
\bibitem{bouck08} D. Bouck, A. Joglekar, and K. Bloom, Annu. Rev. Genet.~\textbf{42}, 335 (2008).
\bibitem{karsenti01} E. Karsenti and I. Vernos, Science~\textbf{294}, 543 (2001).
\bibitem{fuller08} B.G. Fuller, M.A. Lampson, E.A. Foley, S. Rosasco-Nitcher, K.V. Le, P. Tobelmann, 
D.L. Brautigan, P.T. Stukenberg, and T.M. Kapoor, Nature~\textbf{453}, 1132 (2008).
\bibitem{cimini06} D. Cimini, X. Wan, C.B. Hirel, and E.D. Salmon, Curr. Biol.~\textbf{16}, 1711 (2006).
\bibitem{stumpff08} J. Stumpff, G.V. Dassow, M. Wagenbach, C. Asbury, and L. Wordeman, 
Dev. Cell~\textbf{14}, 252 (2008).
\bibitem{greenan10} G. Greenan, C.P. Brangwynne, S. Jaensch, J. Gharakhani, F. J\"ulicher, and A.A. Hyman, 
Curr. Biol.~\textbf{20}, 353 (2010).
\bibitem{varga06} V. Varga, J. Helenius, K. Tanaka, A.A. Hyman, T.U. Tanaka, and J. Howard, Nat. Cell Biol.~\textbf{8}, 957 (2006).
\bibitem{howard07} J. Howard and A.A. Hyman, Curr. Opin. Cell Biol.~\textbf{19}, 31 (2007).
\bibitem{varga09} V. Varga, C. Leduc, V. Bormuth, S. Diez, and J. Howard, Cell~\textbf{138}, 1174 (2009).
\bibitem{johann12} D. Johann, C. Erlenk\"amper, and K. Kruse, Phys. Rev. Lett.~\textbf{108}, 258103 (2012).
\bibitem{melbinger12} A. Melbinger, L. Reese, and E. Frey, Phys. Rev. Lett.~\textbf{108}, 258104 (2012).
\bibitem{masuda87} H. Masuda, W. Z. Cande, Cell~\textbf{49}, 193 (1987).
\bibitem{nedelec02} F. N$\acute{\text e}$d$\acute{\text e}$lec, J. Cell Biol.~\textbf{158}, 6 (2002).
\bibitem{tawada91} K. Tawada and K. Sekimoto, J. Theor. Biol.~\textbf{150}, 193 (1991). 
\bibitem{bieling10} P. Bieling, I.A. Telley, T. Surrey, Cell~\textbf{142}, 420 (2010).
\bibitem{braun11} M. Braun, Z. Lansky, G. Fink, F. Ruhnow, S. Diez, and M.E. Janson, Nat. Cell. Biol.~\textbf{13}, 1259 (2011).
\bibitem{johann14} D. Johann, D. Goswami, and K. Kruse, Phys. Rev. E. {\bf 89}, 042713 (2014).
\bibitem{walcott10} S. Walcott and S. X. Sun, Phys. Rev. E {\bf 82}, 050901(R) (2010).
\bibitem{lansky15} Z. Lansky, M. Braun, A. Ludecke, M. Schlierf, P.R. ten Wolde, M.E. Janson, and 
S. Diez, Cell~\textbf{160}, 1159 (2015).
\bibitem{kapitein08} L.C. Kapitein, M.E. Janson, S.M.J.L. van den Wildenberg, C.C. Hoogenraad, C.F. Schmidt, and 
E.J.G. Peterman, Curr. Biol.~\textbf{18}, 1713 (2008).
\bibitem{valentine06} M.T. Valentine, P.M. Fordyce, T.C. Krzysiak, S.P. Gilbert, and S.M. Block, Nat. Cell Biol.~\textbf{8}, 470 (2006).
\bibitem{valentine06a} M.T. Valentine, P.M. Fordyce, and S.M. Block, Cell Div. {\bf 1}, 31 (2006).
\bibitem{krug91} J. Krug, Phys. Rev. Lett.~\textbf{67}, 1882 (1991).
\bibitem{kruse02} K. Kruse and K. Sekimoto, Phys. Rev. E {\bf 66}, 031904 (2002).
\bibitem{gao15} T. Gao, R. Blackwell, M.A. Glaser, M.D. Betterton, M.J. Shelley, Phys. Rev. Lett.~\textbf{114}, 048101 (2015).

\end{thebibliography}
\end{document}